\documentclass[sigconf]{acmart}

\AtBeginDocument{%
  }

\copyrightyear{2025}
\acmYear{2025}
\setcopyright{cc}
\setcctype{by}
\acmConference[IDC '25]{Interaction Design and Children}{June 23--26, 2025}{Reykjavik, Iceland}
\acmBooktitle{Interaction Design and Children (IDC '25), June 23--26, 2025, Reykjavik, Iceland}\acmDOI{10.1145/3713043.3727057}
\acmISBN{979-8-4007-1473-3/2025/06}

\begin{document}

\title[Participatory Design of GenAI with High School Students]{``How can we learn and use AI at the same time?'': Participatory Design of GenAI with High School Students}


\author{Isabella Pu}
\orcid{0009-0009-0798-0949}
\affiliation{%
  \institution{Massachusetts Institute of Technology}
  \city{Cambridge}
  \state{Massachusetts}
  \country{USA}
}
\email{ipu@media.mit.edu}

\author{Prerna Ravi}
\affiliation{%
  \institution{Massachusetts Institute of Technology}
  \city{Cambridge}
  \state{Massachusetts}
  \country{USA}
}
\email{prernar@mit.edu}

\author{Linh Dieu Dinh}
\affiliation{%
  \institution{Wellesley College}
  \city{Wellesley}
  \state{Massachusetts}
  \country{USA}
}
\email{ld105@wellesley.edu}

\author{Chelsea Joe}
\affiliation{%
  \institution{Dartmouth College}
  \city{Hanover}
  \state{New Hampshire}
  \country{USA}
}
\email{chelsea.c.joe.25@dartmouth.edu}

\author{Caitlin Ogoe}
\affiliation{%
  \institution{Massachusetts Institute of Technology}
  \city{Cambridge}
  \state{Massachusetts}
  \country{USA}
}
\email{ogoe@mit.edu}

\author{Zixuan Li}
\affiliation{%
  \institution{Massachusetts Institute of Technology}
  \city{Cambridge}
  \state{Massachusetts}
  \country{USA}
}
\email{zixuanl4@mit.edu}

\author{Cynthia Breazeal}
\affiliation{%
  \institution{Massachusetts Institute of Technology}
  \city{Cambridge}
  \state{Massachusetts}
  \country{USA}
}
\email{cynthiab@media.mit.edu}

\author{Anastasia K. Ostrowski}
\affiliation{%
  \institution{Purdue University}
  \city{West Lafayette}
  \state{Indiana}
  \country{USA}
} 
\email{akostrow@purdue.edu}

\renewcommand{\shortauthors}{Pu et al.}

\begin{abstract}
As generative AI (GenAI) emerges as a transformative force, clear understanding of high school students' perspectives is essential for GenAI's meaningful integration in high school environments. In this work, we draw insights from a participatory design workshop where we engaged 17 high school students---a group rarely involved in prior research in this area---through the design of novel GenAI tools and school policies addressing their key concerns. Students identified challenges and developed solutions outlining their ideal features in GenAI tools, appropriate school use, and regulations. These centered around the problem spaces of combating bias \& misinformation, tackling crime \& plagiarism, preventing over-reliance on AI, and handling false accusations of academic dishonesty. Building on our participants' underrepresented perspectives, we propose new guidelines targeted at educational technology designers for development of GenAI technologies in high schools. We also argue for further incorporation of student voices in development of AI policies in their schools.
\end{abstract}




\begin{teaserfigure}
  \includegraphics[width=\textwidth]{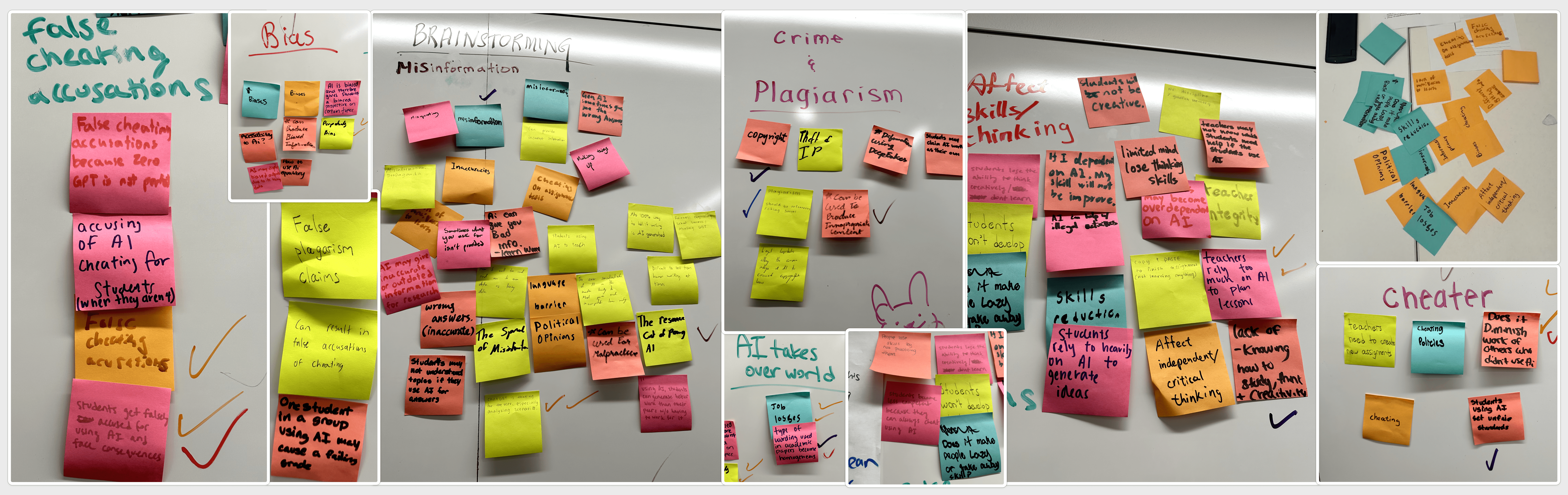}
  \caption{A collection of brainstormed ideas from students in the participatory design workshop. Students were asked to brainstorm concerns they have regarding GenAI in education and group them on the whiteboard.}
  \Description{A collection of sticky notes from student participants that list different problems with GenAI in education in different grouped categories. The categories are False Cheating Accusations, Bias, Misinformation, Crime \& Plagiarism, AI Takes Over World, Affect Skills/Thinking, and Cheater.}
  \label{fig:teaser}
\end{teaserfigure}


\maketitle

\section{INTRODUCTION}
Generative AI (GenAI) is an emerging technology in various fields, including education. ChatGPT, one of the first readily available GenAI tools, even achieved the title of fastest-growing consumer application ever \cite{hu2023chatgpt}. However, its adoption in school settings has been slower, as both teachers and students face a steep learning curve \cite{rashid2024generative}. While GenAI presents challenges like misinformation \cite{walczak2023challenges,solyst2024children}, over-reliance \cite{buccinca2021trust,zhai2024effects}, cheating \cite{ventayen2023chatgpt,waltzer2023students}, and issues of data privacy and bias \cite{huangarticle, salazar2024generative, van2024challenging}, it also offers many benefits such as personalized learning, improved student engagement, rapid feedback, and assistance with grading and lesson planning \cite{maghsudi2021personalized, noroozi2024generative}.

As GenAI becomes increasingly prevalent in education, we face a dual imperative: managing its integration while ensuring foundational skills remain centered \cite{holmes2023guidance}. Students need AI proficiency both to enhance their learning experience \cite{zafari2022artificial} and prepare for future careers \cite{mesko2023prompt, antonenko2023service}, but they must simultaneously be aware of the dangers of inaccurate AI-generated content \cite{walczak2023challenges,solyst2024children} and over-reliance \cite{buccinca2021trust, zhai2024effects}. This demonstrates a fundamental paradox: while AI interaction enables more autonomous learning, over-reliance on these tools risks eroding the very skills they aim to enhance \cite{xie2024artificial}. This is particularly crucial in high school, where students are developing their critical thinking skills \cite{paul1989critical, schafersman1991introduction} in preparation to transition to higher education \cite{venezia2013transitions} or the workforce \cite{blustein2000school}.

Recent guidelines from UNESCO \cite{holmes2023guidance} emphasize the importance of human agency, inclusion, and ethical AI use, while the U.S. Department of Education's guidelines for developers \cite{doe_edtech_ai} highlight the need for trust, safety, and transparency. Both frameworks additionally present the importance of equity. While these guidelines provide valuable high-level principles, they lack specific, actionable steps for real classroom implementation. For these types of guidelines to be truly effective, they must be actionable, clear, and inclusive, involving all stakeholders to ensure thoroughness and fairness \cite{sanusi2024stakeholders}. While students are often the primary users of educational technology (EdTech) tools \cite{zheng_charting_2024}, they are frequently excluded from design discussions and policies surrounding them. K-12 students, in particular, are less commonly involved in EdTech participatory design research \cite{sarmiento2022participatory}, and most K-12 participatory design research focuses on younger students \cite{irgens2022designing,newman_i_2024,buddemeyer2022unwritten}. Including student voices yields valuable insights and helps to better synchronize technology development with their actual needs \cite{jacobsen2013technology,zheng_charting_2024}, yet current frameworks rarely incorporate direct \textit{student} input.

This paper explores participatory design \cite{bjorgvinsson2010participatory, tuhkala2021systematic} involving high school students ($N=17$) in a workshop where they designed GenAI tools and school policies to address key problems they identified. This helped students explore their interests, concerns, {and goals regarding} GenAI. Our research questions are:
\begin{itemize}
    \item \textbf{RQ1:} How can GenAI tools and school policies, designed \textit{with} high school students, address the needs and challenges of high school environments?
    \item \textbf{RQ2:} What design guidelines for GenAI can educational technology designers follow to develop effective and user-centered technologies for high schools?
\end{itemize}

This work bridges the gap between theoretical design frameworks and practical development of GenAI for high school education. We contribute to discussions on participatory design and GenAI in education in three key ways. First, we \textbf{provide insights} into high school students' perspectives through a participatory design workshop. Second, we propose \textbf{actionable guidelines for EdTech designers} that build upon student perspectives and existing frameworks \cite{holmes2023guidance, doe_edtech_ai}. Finally, we demonstrate the importance of \textbf{incorporating student voices in school AI policy development}.

\section{RELATED WORK}
\subsection{The Role of Generative AI in Education}

GenAI holds transformative powers for education, but generally remains in the developmental stages in classrooms due to technological complexities \cite{wang2024critical, ma2024factors}. Many schools hesitate to integrate GenAI, as seen in initial resistance toward and banning of ChatGPT in public schools in New York City and Australia \cite{yu2023reflection,baidoo2023education}. This hesitation stems from various challenges that need careful management in schools \cite{lee2024impact}. One prominent risk is academic dishonesty, as GenAI has performed surprisingly well in subjects ranging from English \cite{de2023can} to law \cite{choi2021chatgpt,ryznar2022exams} and medicine \cite{kung2023performance,fijavcko2023can}. Additionally, current AI detectors cannot reliably identify GenAI-assisted work \cite{chaka2023detecting, oravec2022ai}, complicating efforts to maintain academic integrity while still allowing beneficial AI use.

Young learners, still developing critical thinking skills \cite{paul1989critical, schafersman1991introduction}, are particularly vulnerable to accepting AI-generated content without scrutiny, including errors called ``hallucinations'' \cite{walczak2023challenges} or misinformation arising from limitations of training data \cite{mccausland2020bad}. This is especially problematic for education \cite{holmes2023guidance}, particularly when AI perpetuates existing biases by generating content that discriminates against marginalized groups \cite{salazar2024generative} or reinforces stereotypes \cite{van2024challenging}. To mitigate this, GenAI integration must be inclusive and specifically address accuracy and equity concerns \cite{francis2025generative}. Additional challenges include privacy issues risking unauthorized access to personal data \cite{huangarticle} and equitable access, as disparities in school funding \cite{liu2024digitalequity} and home technology use \cite{thomas2022closing} could limit GenAI access and further widen achievement gaps \cite{van2008technology}. 

Additionally, as AI extends beyond education, proficiency in its use becomes increasingly vital for future careers \cite{mesko2023prompt, antonenko2023service}. More AI experience can help students prepare for a rapidly evolving technological workplace while mitigating classroom risks \cite{ali2021children, lee2021developing}. Students should {also} continue to develop their foundational skills, as dependence on AI could diminish social engagement and creativity, as well as lead to over-reliance on AI generated information \cite{buccinca2021trust, zhai2024effects}. Though some advocate for banning GenAI entirely \cite{lau2023ban}, many recognize its value in learning \cite{konecki2024teachers}, as GenAI improves classroom efficiency \cite{ayanwale2022teachers} and enables personalized interactions \cite{holmes2023guidance,noroozi2024generative} that enhance student learning. GenAI tools can adapt to individual students' needs---especially helpful for English as an Additional Language students and students with special needs \cite{liu2023understanding}---enabling more meaningful classroom interactions \cite{lee2024impact, maghsudi2021personalized} and boosting student engagement and motivation through interactive, creative activities \cite{nguyen2024enhancing, huang2023effects}. 

Our study addresses key gaps in understanding the impact of GenAI on high school students. Positioned at a critical transitional phase, these students must simultaneously develop AI literacy and preserve core skills to navigate future education and careers \cite{venezia2013transitions, blustein2000school}. While prior research has addressed general benefits and risks of GenAI, limited attention has been given to the unique needs and experiences of high school students. Though challenges posed by AI use, like hallucinations, are well-documented, their impact on high school students' perceptions of GenAI remains underexplored. We use students' perspectives as a lens to investigate how GenAI can enhance learning outcomes while fostering equity, critical thinking, and creativity. 

\subsection{Current GenAI Policies for EdTech Designers \& Developers}

Design of new GenAI for education remains constrained by a lack of comprehensive, actionable guidelines for developers \cite{holmes2020artificial} and a disconnect between designers versus educators and students \cite{luckin2019designing}. While guidelines from UNESCO \cite{holmes2023guidance} and the U.S. Department of Education (DoE) \cite{doe_edtech_ai} provide valuable principles for AI EdTech designers, they fall short of proposing practical, actionable strategies that developers need to effectively navigate the complexities of educational technology.

Both UNESCO’s \cite{holmes2023guidance} and the DoE’s \cite{doe_edtech_ai} guidelines emphasize a human-centered approach to generative AI, prioritizing equity, inclusion, and human agency. They address risks such as data privacy breaches, algorithmic bias, and threats to academic integrity while advocating for ethical validation of AI systems and equitable access to AI tools. Both highlight the importance of ensuring tools are inclusive, accessible, and focused on mitigating bias to support all learners equitably. Additionally, they illustrate the need for transparency and trust-building to ensure AI tools align with educational values and protect civil rights. While both guidelines include some recommendations for designers, UNESCO’s framework contains only a few relevant sections, whereas the DoE’s guidelines are explicitly tailored for EdTech developers.

While other frameworks exist for responsible AI development in general, such as the National Institute of Standards and Technology AI Risk Management Framework \cite{nist_ai2024artificial} and the Microsoft Responsible AI Standard \cite{microsoft_responsible_ai}, these are not directed toward EdTech developers. Similarly, ethical guidelines focused on child AI use---such as UNICEF’s Policy Guidance on AI for Children \cite{unicef_children}---highlight critical protections for minors but are not targeted at developers or specific to education. Teacher- and student-facing guidelines also exist \cite{miao2024aistudents,miao2024aiteachers}, but again, these are not targeted towards developers. To our knowledge, the guidelines from UNESCO and the DoE are the only ones---at the time of writing---that are directed (or have sections directed) to developers of \textit{educational} \textit{AI} tools.

Compounding the scarcity of detailed guidelines for AI EdTech designers \cite{holmes2020artificial} is the tendency for developers to work in isolation from schools and educators, often lacking expertise in learning practices and missing student and teacher perspectives \cite{luckin2019designing}. As a result, AI tools often overlook practical classroom challenges and the nuanced needs of classrooms \cite{ayeni2024ai}. While UNESCO and DoE guidelines recommend co-design practices, neither was co-designed with students, limiting their ability to address student needs. Additionally, these guidelines do not focus on a specific age range, whereas our focus on high school students enables targeted insights, and bridging this gap requires a participatory approach involving stakeholders \cite{sarmiento2022participatory, sanusi2024stakeholders}. Through our participatory design approach, we are able to incorporate valuable high school perspectives to better direct tool development with actionable guidelines.


\subsection{Participatory Design of Technology for Education}

Traditional educational technology development often lacks direct user feedback, resulting in tools that face resistance from both teachers and students \cite{kohn1993choices}. Participatory Design (PD), a method that directly involves those affected by a design change in the design process \cite{bjorgvinsson2010participatory,tuhkala2021systematic}, addresses these issues through engaging stakeholders and leads to tools that are more functional and aligned with user needs \cite{jacobsen2013technology,zheng_charting_2024,triantafyllakos2008we}. As PD emphasizes designing \textit{with} users, centering their values and empowering them to make changes that coincide with their goals \cite{spinuzzi_methodology_2005, correia2008don}, PD can foster a sense of ownership, agency, and authority in students to shape their educational environment \cite{booker_participatory_2016}. Engaging students in PD deepens their connection to the tools they use \cite{levin2000putting,dindler2020computational} while fostering critical thinking, collaboration, problem-solving \cite{bell2016learning,wu_investigating_2021}, and social skills like cooperation and respect \cite{triantafyllakos2008we}, preparing them for a technology-driven world \cite{bodker_participatory_2018}. Researchers have also leveraged PD to engage students and teachers in shaping ethical frameworks for AI in education \cite{alfredo2024human, brossi2022student}. This demystifies the AI ``black box'' and presents AI to users in more understandable ways \cite{khosravi2022explainable}. Finally, PD can reshape power dynamics and foster a culture of care, driving systemic change in EdTech and the broader sociocultural context of schools \cite{wake2013developing, higgins2019power}.

In the context of GenAI for education, PD has helped develop user-centered AI, such as conversational agents and tools for expressing creativity \cite{newman_i_2024,buddemeyer2022unwritten}. However, high school students have often been excluded from recent PD, including efforts on AI and other technology \cite{sarmiento2022participatory}---recent EdTech PD initiatives have mainly involved teachers \cite{gardner2022co,lin2021engaging,cober_teachers_2015}, higher education students \cite{zheng_charting_2024,gros2016students}, or younger students \cite{irgens2022designing,newman_i_2024,buddemeyer2022unwritten}. 
Our work directly addresses this gap by engaging high school students as active participants designing GenAI tools and accompanying school policies. 
By using PD to connect GenAI EdTech design with the lived realities of high school students, we aim to ensure resulting technologies are effective and empower students to take ownership of their learning \cite{Gallup_2019}.

\section{METHODS}

\begin{table*}[ht!]
  \caption{Overview of students’ demographics (Note: Percentages may not add up to 100\% due to rounding.)}
  \label{tab:student-demographics}
  \centering
  \begin{tabular}{|c|c|c|}
    \toprule
     \textbf{Demographic} & \textbf{\# of Students} & \textbf{\% of Students} \\
    \midrule
    \hline
     \multicolumn{3}{|c|}{\textbf{\textit{Gender}}} \\
    \hline 
     Female & 7 & 41.2\% \\
    \hline 
     Male & 10 & 58.8\% \\
    \hline 
     \multicolumn{3}{|c|}{\textbf{\textit{Grade}}} \\
    \hline 
     9 & 2 & 11.8\% \\
    \hline 
     10 & 1 & 5.9\% \\
    \hline 
     11 & 7 & 41.2\% \\
    \hline 
     12 & 7 & 41.2\% \\
    \hline 
     \multicolumn{3}{|c|}{\textbf{\textit{Age}}} \\
    \hline 
     14 & 2 & 11.8\% \\
    \hline 
     15 & 2 & 11.8\% \\
    \hline 
     16 & 7 & 41.2\% \\
    \hline 
     17 & 6 & 35.3\% \\
    \hline 
     \multicolumn{3}{|c|}{\textbf{\textit{Ethnicity}}} \\
    \hline 
     American Indian or Alaska Native & 1 & 5.9\% \\
    \hline 
     Asian & 10 & 58.8\% \\
    \hline 
     Black or African-American & 3 & 17.6\% \\
    \hline 
     White & 3 & 17.6\% \\
    \hline 
     \multicolumn{3}{|c|}{\textbf{\textit{School Type}}} \\
    \hline 
     Public & 7 & 41.2\% \\
    \hline 
     Private & 10 & 58.8\% \\
    \hline 
     \multicolumn{3}{|c|}{\textbf{\textit{Previous AI Experience}}} \\
    \hline 
     Text Generation (i.e. ChatGPT) & 15 & 88.2\% \\
    \hline 
     Image Generation (i.e. Midjourney) & 3 & 17.6\% \\
    \hline 
     AI Integrated in Other Apps (i.e. Adobe Firefly) & 6 & 35.3\% \\
    \hline 
     Other AI Tool & 1 & 5.9\% \\
    \hline 
     No Experience & 1 & 5.9\% \\
    \hline
  \end{tabular}
  \Description{Table describing demographics of the 17 students who participated in the workshop, in aggregate. The demographics are Gender, Grade, Age, Ethnicity, School Type, and Previous AI Experience.}
\end{table*}

Our \textbf{participatory design workshop with high school students} was part of a broader study that also included interviews with seven high school teachers. This paper focuses specifically on the workshop, which was shaped by themes that emerged from the teacher interviews, such as concerns about AI-related privacy, the slow pace of school policy development, and the lack of professional development opportunities. The workshop itself included both AI tool design and school policy design components.

All work in this study was approved by the university’s Institutional Review Board. 

\subsection{Participants}

We recruited 17 high school students (demographics in Table \ref{tab:student-demographics}) via mailing lists of students interested in AI or  robotics. All students assented and had parental consent. Participants, aged 14 to 17 (average age 16) were from a mix of public ($N=7$) and private schools ($N=10$). 16 students had prior experience with GenAI, with most having used text generation tools like ChatGPT ($N=15$). Each student was compensated with a \$50 Amazon gift card for participating.

\subsection{Workshop Structure}

\begin{figure*}[ht!]
\includegraphics[width=0.98\textwidth]{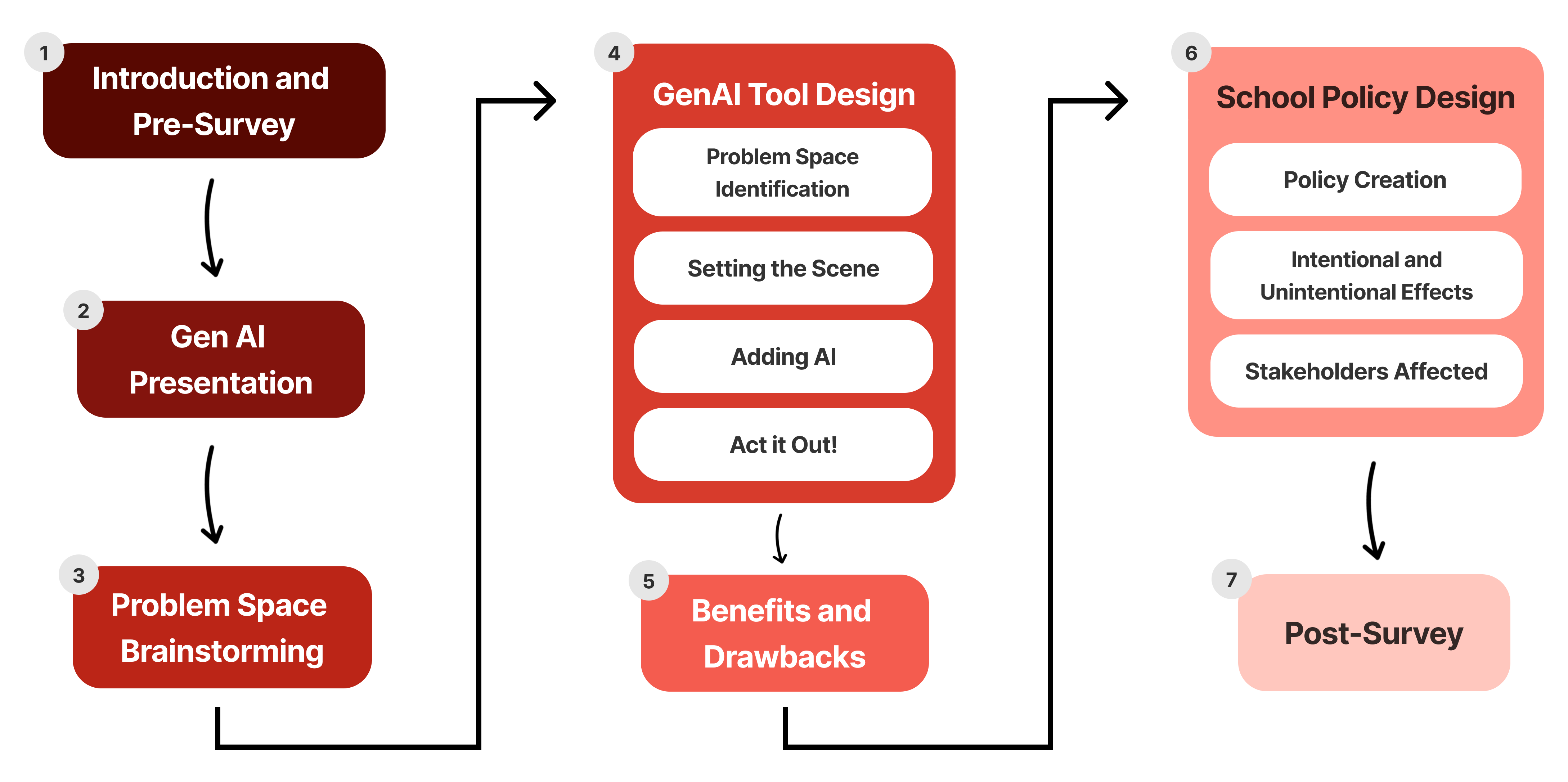}
  \caption{Diagram depicting the structure of our participatory design workshop with high school students.}
  \Description{A diagram depicting the structure of the student workshop, with 7 steps: (1) Introduction and Pre-Survey, (2) Gen AI Presentation, (3) Problem Space Brainstorming, (4) GenAI Tool Design, which has four worksheets that are completed during it: Problem Space Identification, Setting the Scene, Adding AI, and Act it Out, (5) Benefits and Drawbacks, (6) School Policy Design, which has three activities completed during it: Policy Creation, Intentional and Unintentional Effects, and Stakeholders Affected, (7) Post-Survey.
  }
  \label{fig:workshop}
\end{figure*}

\hfill \break
\indent Our 7-step workshop structure is in Figure \ref{fig:workshop}.

\paragraph{Introduction \& Pre-Survey:} The workshop began with a brief introduction to the topic, icebreakers, and a pre-survey. The pre-survey measured participants’ previous experiences and confidence with GenAI tools and assessed student perspectives on their schools’ current AI use (full survey available in supplementary materials).

\paragraph{GenAI Presentation:} We presented an overview of GenAI to provide a unified platform for workshop participation, covering: example uses (art, music, video, text, and code generation), chatbots, search engines vs. GenAI, how ChatGPT works (next word prediction) and training data, a brainstorming activity where students identified GenAI's \textit{current uses} in education, a brainstorming activity about benefits and risks, and strategies to explore GenAI biases. {Concepts included were informed by existing AI education materials \cite{hollands2024establishing, lee2021developing, ali2021children}.} The presentation equipped participants with the knowledge and vocabulary to engage confidently in discussions \cite{sanders2008co}, regardless of prior experience, {and used} playful engagement to help them visualize potential futures \cite{vaajakallio_design_2014}.

\paragraph{Problem Space Brainstorming:} Post-presentation, students individually brainstormed \textit{concerns} about GenAI in education, covering areas like homework, classwork, and testing. They wrote these concerns on sticky notes (seen in Figure \ref{fig:teaser}) and posted them on a whiteboard for group sharing. Students used affinity diagramming \cite{harboe2015real} to group these concerns into six thematic domains. Facilitators were present but avoided guiding students or introducing specific issues, ensuring that identified concerns were based entirely on students' perspectives.

\paragraph{GenAI Tool Design:} Based on the thematic domains, participants formed four small groups based on the problem space they found most interesting. Each group was tasked with generating a specific education-related problem from their chosen domain, envisioning a real-world scenario where a GenAI tool could be applied, then designing said tool. Students completed four worksheets that guided them through conceptualizing their GenAI tools (two worksheets highlighted in Figure \ref{fig:worksheets} and example output in Figure \ref{fig:example-and-photo}(a); full worksheets in supplementary materials).

\begin{figure*}[ht!]
  \includegraphics[width=0.95\textwidth]{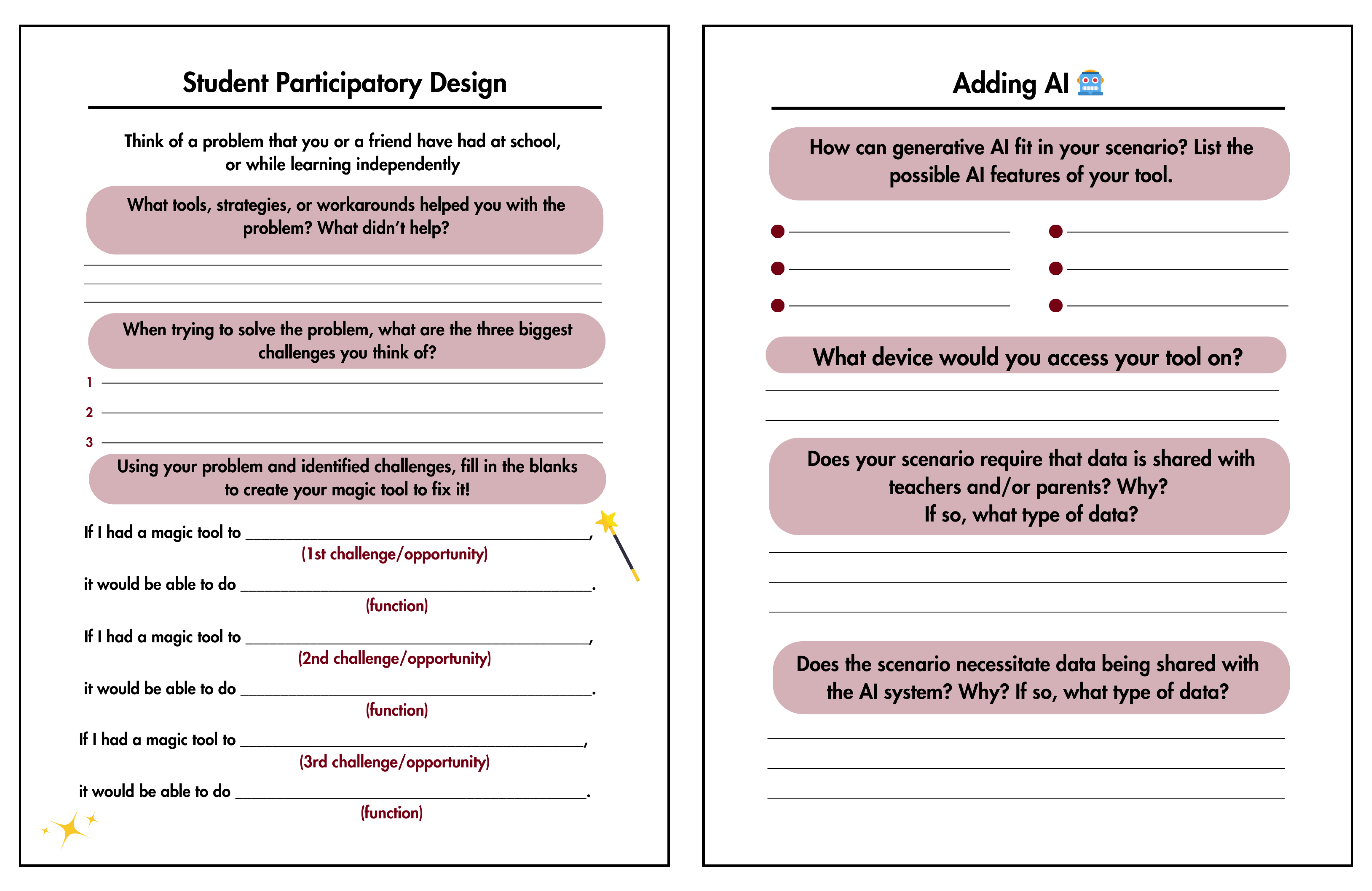}
  \caption{Spotlight on two of the four worksheets (Left: Problem Space Identification; Right: Adding AI) that guide students through the GenAI Tool Design portion of the workshop.}
  \Description{A diagram depicting two of the four worksheets guiding student AI tool design. On the left: Problem Space Identification, on the right: Adding AI.
  
  Problem Space Identification asks students to first think of a problem they have had in education, then think about tools, strategies, workarounds, and challenges involved in trying to fix the problem. Finally it asks students, if they had a magic tool that could do anything to solve the problem, what would it do? Adding AI guides students in how generative AI could power their magic tool, as well as asks students what device their tool is accessed on and whether they scenario necessitates data being shared.}
  \label{fig:worksheets}
\end{figure*}

In the first worksheet, \textbf{``Problem Space Identification''}, students identified a specific problem within their chosen domain to solve using GenAI. This exercise helped them understand existing solutions and challenges in their chosen space, ending with brainstorming three new ways to address the problem.

Next, in \textbf{``Setting the Scene''}, participants ideated a scenario featuring their GenAI tool. Groups considered factors including stakeholders, how the tool would fit into workflows, and the actions of characters in their scenario.

In \textbf{``Adding AI''}, groups specified how GenAI would be used in their tool. They detailed the tool’s features, user interface, data processing, content generation, user feedback, data access, and device compatibility.

In \textbf{``Act it Out!''}, each group wrote a 3-minute skit to demonstrate their tool in action, describing setting, characters, and interactions. Role-playing was used to help students understand the tool’s impact on users and others \cite{stromberg2004interactive,iacucci2000move}. Students reflected on observed patterns, new concerns, and how the process shaped their views on AI through a post-skit discussion.

\paragraph{GenAI Tool Benefits and Drawbacks:} As a large group, students discussed the benefits and consequences of using the GenAI tools they designed, various stakeholders, and intentional versus unintentional side effects.

\paragraph{School Policy Design:} In the same small groups as the GenAI Tool Design section, students developed school policies to address specific drawbacks of GenAI that emerged during group discussions. Groups had the flexibility to focus on any identified drawback rather than being limited to issues from their own tool. To scaffold, we provided a structured example of a non-AI-related school policy, highlighting key components: goals, stakeholders, and impacts. Each group then developed a detailed policy addressing one specific drawback, outlining potential benefits, harms, and affected stakeholders. These policies were shared at the end of the workshop.

\paragraph{Post-Survey:} Students completed a post-survey (full survey in supplementary materials) with questions similar to the pre-survey, assessing familiarity with AI, confidence in using it, and views on school AI policies. It also explored changes in opinions on AI in education, interest in future GenAI tool or policy development, and key takeaways. For the purpose of this paper, we focus on components regarding students' goals for future engagement with GenAI in schools, any changes in feelings about GenAI, and key takeaways.

\begin{figure*}[t!]
\includegraphics[width=0.98\textwidth]{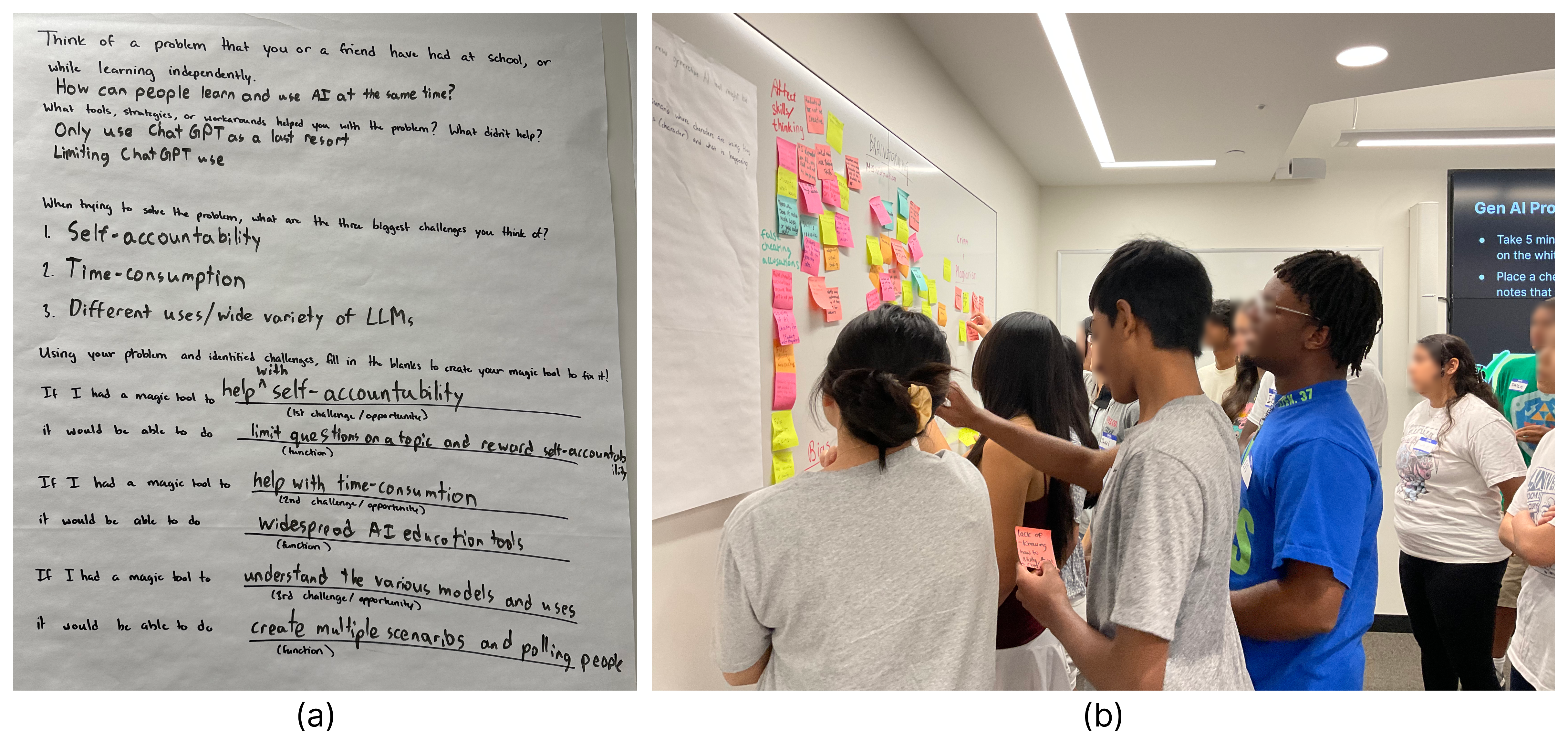}
  \caption{(a) An example worksheet output filled out by student participants. (b) Students engaging with the workshop activities.}
  \Description{
  On the left, in sub-figure (a), is a photograph of a Problem Space Identification worksheet filled out by student participants of the participatory design workshop. This was filled out by the group investigating over-reliance on AI and contains a problem they want to solve (How can people learn and use AI at the same time?), tools or workarounds that have helped with the problem (only using ChatGPT as a last resort, limiting ChatGPT use), and three challenges in solving the problem (self-accountability, time consumption, and the wide variety of LLMs). Finally, students filled it out with three examples of magic tools that could help tackle those challenges. The first was a tool that could limit questions on a topic and reward self-accountability, the second was using widespread AI education to help with time consumption, and the third was creating multiple scenarios and polling people to better understand the various AI models and uses.

  On the right, in sub-figure (b), a group of high school students of different backgrounds is standing in front of a whiteboard full of colorful sticky notes placed in groups. One student in the center is holding a sticky note that says "Lack of knowing how to study" and considering where to place it on the whiteboard.
  }
  \label{fig:example-and-photo}
\end{figure*}

\subsection{Data Collection}

To document both small- and full-group discussions, audio recorders were placed at each participant table. The audio recordings were transcribed using automated AI services, then manually verified by researchers for accuracy. All artifacts produced during the workshop (e.g., brainstorming sticky notes, worksheets, etc.) were digitally photographed and transcribed for analysis.

\subsection{Data Analysis}

We employed an inductive qualitative approach \cite{guest_applied_2012} to analyze the workshop data. Four researchers independently reviewed the workshop transcripts to identify emergent themes. After finalizing the list of themes collaboratively, two independent researchers coded each transcript, resolving any discrepancies through discussion \cite{daly_design_2012,ostrowski2022mixed}.

Relevant questions from our surveys were either multiple-select or free response. We analyzed multiple-select questions by counting and graphing each selection. Free-response questions were {qualitatively} reviewed by two researchers who independently identified themes, finalized them collaboratively, and then coded each response, resolving any discrepancies \cite{daly_design_2012,ostrowski2022mixed}.

\section{FINDINGS}

Three main themes emerged from the student workshop, including both tool and policy design discussions: {\textbf{AI Tool Features} (4.1)}, referring to when students brought up new features they wished to see or concerns they had about existing features of AI tools; {\textbf{School/Classroom Use} (4.2)}, referring to how they wish to see AI used and current challenges of AI use in the classroom; and {\textbf{Regulations} (4.3)}, referring to their desire for more regulation and their questions about current regulations.

Themes and sub-themes, with definitions, are located in Table \ref{tab:workshop-codes}. Relevant sub-themes are also noted in-text with parentheses. In this section, we examine challenges and needs identified by students per theme, as well as proposed designs. Where students were unable to propose solutions, we highlight unresolved questions and concerns. These insights collectively address RQ1. Sections 4.1 to 4.3 present workshop outcomes (by theme), incorporating both verbal and written contributions. Section 4.4 presents student-created school policies and section 4.5 presents survey results.

\begin{table*}[ht!]
\caption{Overview of themes and sub-themes found in transcripts of student discussions in participatory design workshop}
\label{tab:workshop-codes}
\centering
\resizebox{\linewidth}{!}{
\begin{tabular}{l|l|l}
\hline
\textbf{Themes}& \textbf{Sub-themes}& \textbf{Definitions}                                                          \\ \hline
{AI Tool {Features}}& Fair and Equitable AI Use& Ensuring AI tools are used in a fair and equitable manner for all students.\\
                                    & AI Hallucinations& Concerns about AI generating incorrect or misleading information.\\
                                    & AI Should Cite Its Sources& Requiring AI to provide citations for the information it generates.\\
                                    & AI Bias Concerns& Concerns about AI producing biased or discriminatory outputs.\\
 & Quality of Training Data&Ensuring AI is trained on high-quality and diverse datasets.\\
                                     \hline
School/Classroom {Use}& Personalized Learning& Using AI to tailor educational content to individual student needs.\\
 & Balancing AI \& Learning&Finding the right mix between AI usage and traditional learning methods.\\
 & Lack of Teacher AI Literacy&Concerns about teachers' insufficient knowledge or skills regarding AI.\\
 & Academic Integrity&Ensuring AI use does not encourage cheating or academic dishonesty.\\
                                    & Unreliable AI Detectors& Concerns about the accuracy and fairness of AI detection tools.\\
                                    \hline
 Regulat{ions} & AI Needs More Regulation&Advocating for stricter oversight and regulation of AI tools in education.\\
 & The Roles of Teachers \& Parents&Questions about the roles of educators and parents in AI use.\\
 & Access to Data, With Limitations&Allowing AI access to private data with certain rules in place.\\
 \hline
\end{tabular}}
    \Description{The table presents themes, sub-themes, and definitions for each sub-theme from the student workshop.

    The first theme was AI Tool Features, with the following sub-themes: Fair and Equitable AI Use (Ensuring AI tools are used in a fair and equitable manner for all students), and AI Hallucinations (Concerns about AI generating incorrect or misleading information), and AI Should Cite Its Sources (Requiring AI to provide citations for the information it generates), and AI Bias Concerns (Concerns about AI producing biased or discriminatory outputs), and Quality of Training Data (Ensuring AI is trained on high-quality and diverse datasets).

    The second theme was School/Classroom Use, with the following sub-themes: Personalized Learning (Using AI to tailor educational content to individual student needs), and Balancing AI \& Learning (Finding the right mix between AI usage and traditional learning methods), and Lack of Teacher AI Literacy (Concerns about teachers’ insufficient knowledge or skills regarding AI), and Academic Integrity (Ensuring AI use does not encourage cheating or academic dishonesty), and Unreliable AI Detectors (Concerns about the accuracy and fairness of AI detection tools).

    The third theme was Regulations, with the following sub-themes: AI Needs More Regulation (Advocating for stricter oversight and regulation of AI tools in education), and The Roles of Teachers \& Parents (Questions about the roles of educators and parents in AI use), and Access to Data, With Limitations (Allowing AI access to private data with certain rules in place).
    }
\end{table*}

Six problem spaces were identified during brainstorming: Over-reliance on AI, Cheating Accusations, Bias, Misinformation, Crime \& Plagiarism, and AI Takes Over the World. Students combined Bias and Misinformation because they recognized that biased AI is likely to produce misinformation, and both issues can stem from problems with training data. Four final student groups {(G)} emerged: \textit{Bias \& Misinformation {(G1)}}, \textit{Crime \& Plagiarism {(G2)}}, \textit{Over-reliance on AI {(G3)}}, and \textit{Cheating Accusations {(G4)}}. We reference groups by their abbreviations below.

\subsection{AI Tool Features}

Throughout the workshop, students looked to minimize bias and misinformation {(AI Bias Concerns)} in AI. An early challenge was inherent bias in human-sourced data: \textit{``One of the biggest problems with bias is that we’re all inherently biased'' (G1)}. Students then identified ways they felt they were inherently biased, eventually concluding \textit{``We're biased based on where we live or how we grew up. AI could be trained by a person of certain beliefs, and that could dictate all its responses'' (G1).} This led to the understanding that because humans will always be involved in producing data and training AI, and humans are inherently biased by their surroundings, AI will always carry some degree of bias.

After recognizing that eliminating bias entirely is impossible, students moved onto methods of addressing it. When brainstorming problems they wanted AI to solve, students mentioned, \textit{``If [AI] were able to make bias, it should be able to fix bias'' (G1)}. A prominent suggestion was that AI should be intentionally exposed to biases during training to learn how to identify and address them: \textit{``If [AI] doesn't ever get exposed to bias, it won't know what bias is. And you can't fix bias without knowing what it is'' (G1).} Building on this, another participant suggested that even if complete bias mitigation is impossible, bias recognition is valuable: \textit{``Even if [the AI] can't fix bias, if it can recognize it, that's okay too'' (G1).} They thought AI systems could flag potential biases by identifying conflicting information or historical patterns that might indicate prejudiced assumptions.

Students also prioritized diverse, accurate training data {(Quality of Training Data)}. With the goal of ``fixing bias'', students recognized that bias \textit{``stems from incorrect data'' (G1)} (Quality of Training Data); hence they suggested ideas such as filtering bias out of training data and improving data diversity so AI is exposed to more unbiased material. They also discussed the challenges of subjective information, such as interpretation of historical events, and agreed that producing nuanced AI outputs would require training AI on diverse perspectives. However, students acknowledged that even with \textit{diverse} data, the \textit{accuracy} of sources remains critical to producing trustworthy outputs.

In this vein, students frequently expressed concerns about AI-generated false information {(AI Hallucinations)}, drawing from personal experiences with AI providing incorrect information. This posed a particular challenge in education. While students appreciated AI's ability to \textit{``congregate data from different sources on the internet'' (G3)}, they worried about AI's inability to evaluate source credibility. Based on their experience with online information---which AI is often trained on---students reflected on the process of distinguishing between reputable and non-reputable sources. They recognized that while some online sources are trustworthy, most are not, and expressed concern that AI systems lack this crucial ability to discern source quality.

\begin{quote} {\textit{``There's a ton of misinformation... a bad source could say, `Here's my opinion,' and say something wrong, and the AI would just learn it. And then it would tell it to you like a fact.'' --G3}} \end{quote}

To address these concerns and leverage existing skills, students proposed several ideas centered around source citation and verification (AI Should Cite Its Sources). They wanted AI to cite its sources, allowing students to verify content authenticity using their academic training. As one student suggested, AI could \textit{``give quotes from the text and ask the student to think for themselves'' (G3)}, indicating students' desire to maintain control over source verification while still benefiting from AI's ability to aggregate content. Students drew parallels to their academic experience and stressed the importance of distinguishing source credibility: \textit{``When you write an essay and you can choose between using a textbook or a blog, you choose the book... it's like that, but for AI'' (G4).}

Beyond individual verification, they also advocated for oversight of the citation process to prevent fabricated sources. These concerns extended to ethical considerations about training data (Quality of Training Data), with students questioning implications of unintentional plagiarism and the ethics of using AI models trained on potentially problematic sources: \textit{``What if the learning set isn’t ethical? Illegally obtained? Is sharing it still ethical?'' (G2)}. This discussion sparked a debate about the dangers of unethically obtained training data. While one student downplayed the impact of obtaining data without explicit consent, another challenged this view by personalizing the issue: \textit{``What if it was your data and information being stolen? You might think it is a bigger deal than when someone else's hard work is being stolen'' (G2).} This exchange led students to conclude that controlling the quality and sourcing of training data was just as crucial as ensuring proper citation in AI outputs.

To mitigate the challenges of hallucinations, G1 proposed an AI browser extension that would enable users to verify AI responses by cross-referencing outputs with trusted sources and providing updated, \textit{reliable} citations {(AI Should Cite Its Sources)}. Their choice of a browser extension was intentional, ensuring the tool could be \textit{``accessed on any device,''} highlighting their interest in accessibility {(Fair and Equitable AI Use)}. The emphasis on accessibility was not unique to G1---\textit{all} groups independently specified that their proposed GenAI tools should be accessible across different devices, reflecting a broader student agreement about the importance of equitable AI access.

\subsection{School/Classroom Use}

In multiple groups, students expressed concerns about AI misinformation spreading through teacher-student relationships. Their primary worry was that teachers, whom students trust implicitly, might unknowingly spread AI-generated misinformation due to limited understanding of the technology (Lack of Teacher AI Literacy). As one student noted during discussions about classroom AI use, \textit{``Teachers don't have full credibility about AI'' (G4)}. This knowledge gap was later attributed to AI's rapid emergence in educational settings, leaving teachers with insufficient resources to develop comprehensive understanding. The concern was heightened by students' reliance on teachers as trusted sources: \textit{``I would call my teacher, and my teacher would give me a valid source... But what if they got a bad source from the AI and didn't know either?'' (G1)}.

To improve teachers' AI literacy, students proposed increased AI training for educators while recognizing practical challenges: \textit{``Maybe a class some teachers would have to take would help... but that's time consuming... I think that would also be an issue for the teachers'' (G3).} They further emphasized that AI training would need to be school-sponsored, compensated, and conducted during school hours, noting \textit{``Teachers and students might lose time in the classroom'' (G3).}

Students also identified a significant challenge in maintaining academic integrity with AI tools, particularly in writing assignments where AI can quickly generate content but lack depth. Their concerns extended beyond cheating to broader issues of educational equity and the diminishing value of hard work:

\begin{quote} {\textit{``With ChatGPT, you can just get it to write stuff for you all the time, and there are ways to make it sound human. It gives good writing and good ideas. And you can get a better grade than others who didn't use it.'' --G4}} \end{quote} 

While AI detectors were a potential solution, students were overwhelmingly critical of teachers’ potential reliance on them, due to their inaccuracy (Unreliable AI Detectors) and stress of false accusations. One noted, \textit{``The problem is that sometimes people write similar to AI just on their own. And that creates a bunch of false accusations'' (G4).} Students were also worried about vulnerable populations: \textit{``[AI detectors] can disproportionately affect international students because of the way they were taught to write'' (G1).} Despite these concerns, students acknowledged why teachers might rely on AI detectors, as current more manual cheating methods---such as re-typing AI-generated text---easily bypass teachers' usual scrutiny of edit history checks.

Students first proposed prioritizing in-class assignments for easier monitoring of AI use but recognized the difficulty of fully preventing AI-assisted cheating: \textit{``People are going to use AI anyway because it's pervasive. It's everywhere in our lives'' (G4).} Recognizing that avoiding AI usage in assignments entirely was unrealistic, students began exploring ways to balance AI features to foster learning and ethical behavior rather than rewarding academic dishonesty. They asked, \textit{``How can people learn and use AI at the same time?'' (G3)}, cautioning against the outsourcing of critical thinking and expressing concern about the erosion of foundational skills (Balancing AI \& Learning). They also linked these concerns to increasing stress among high schoolers, driven by heavy workloads and tight deadlines. Students described their reluctantly growing dependence on AI: \textit{``AI is my last resort, but I have to resort to it because I'm just so pressured'' (G3).} They additionally worried, \textit{``If a huge percent of the population uses ChatGPT for a lot of things, a lot of normal human skills are not going to be used anymore'' (G3)}, giving examples of logic and critical thinking. 

However, students strongly opposed banning GenAI, recognizing its importance for future careers: {\textit{``If we don't use AI now, we'll get to the real world and won't know how to use AI to our advantage'' (G4)}.} Students saw AI as an integral part of the future and that learning to use it effectively was essential, even if they felt hesitant about its growing presence. Instead of eliminating AI, students advocated for designing tools that enhance learning while discouraging over-reliance through design. For example, they suggested having AI only provide incremental hints rather than direct answers. This reflected their desire to integrate AI in a way that maintains academic integrity while encouraging effective learning through methods like personalization (Personalized Learning). Students found value in AI’s ability to address individual needs by focusing on areas of student struggle while encouraging independence in their strengths: 

\begin{quote}
    \textit{``Each person has areas where they don't know things and areas where they know a lot, so it would make more sense for AI to find that out and attack problems based on that'' --G3.}
\end{quote}

Building on this vision, G3 proposed a design for a tool that adapts prompts after being trained on past interactions to better balance learning new material with foundational skill development {(Balancing AI \& Learning)}, such as through logical deduction exercises with provided hints. Students emphasized that this would promote self-accountability and ensure that AI serves as a supportive resource, fostering independence rather than dependency, even in times of increased stress.

\subsection{Regulations}

Students emphasized a need for regulatory oversight (AI Needs More Regulation) but disagreed on an ideal implementation. Although unfamiliar with specific policies, students expressed concerns about current unethical practices, such as unauthorized use of personal data, highlighting a need for stronger safeguards. Some argued in favor of national governance of AI tools: \textit{``We would have to get the government involved to regulate, holding ChatGPT accountable'' (G2).} For globally-used tools, suggestions included forming an international AI committee or adhering to the governing laws of the country the AI company is headquartered in.

On the other hand, some strongly opposed the idea of governmental oversight: \textit{``With the government in control... they can do censorship, they could limit the amount of information that we have'' (G1)}, and preferred regulation through schools, teachers, or parental consent {(The Roles of Teachers \& Parents)}. Students expressing this concern voiced mistrust of government regulation, fearing that issues like AI over-reliance or misinformation could be leveraged by authorities to spread propaganda or limit access to education for specific populations. Students also felt that while parents should be informed about AI tools their children use in school, teachers should have greater control over the tools' outputs and the inputs students provide because \textit{``The teachers are the ones who actually use it'' (G3).}

Students also highlighted the challenge of regulating AI \textit{development}, advocating for laws that hold companies accountable for unethical practices while avoiding penalties for users' unintentional misuse. They were particularly aware that AI tools can output content derived from stolen work:

\begin{quote}
    \textit{``The whole [training] database is probably mostly illegal... what's beneficial about AI regulation is that we can hold companies accountable for stealing. Right now, they can steal anything and it's hard to tell.'' --G2.}
\end{quote}

Students expressed significant concern that stolen content in training data could be unintentionally used by well-meaning users, emphasizing that expecting users to constantly monitor AI-generated content for copyright violations is unrealistic. Although uncertain whether government regulations could fully address the issue---given varying definitions of \textit{``stolen work''} and differing tolerance across countries---they felt that implementing regulations would be a notable improvement over the perceived lack of oversight in the current landscape.

Building on this, two groups designed GenAI tools that centered AI regulation. G2 proposed a tool regulated by both the government and libraries {(AI Needs More Regulation)} to provide verified and sourced information {(AI Should Cite Its Sources)}. Libraries would manage access to the tool but would require government approval to provide it to students, giving the government a role in overseeing its reliability. This approach aimed to ensure legally and ethically sound AI content, protecting against corruption in any single regulatory body: \textit{``Even if the government thinks violating one person's privacy is for the greater good, [the library] can still protect [them]'' (G2).} 

On the other hand, G4 designed a GenAI tool that helped teachers regulate AI, providing detailed reports of students' AI interactions {(The Roles of Teachers \& Parents)}. This approach allowed teachers to oversee and personalize AI use based on their classroom's rules instead of relying on AI detectors. G4 stated:

\begin{quote} {\textit{``There should be teacher control so they can decide which parts students can use... Maybe your teacher will let you use AI for math, but not writing. Or the tool will be more limited for younger students.'' --G4} } \end{quote}

Additionally, students stressed the importance of responsible data management by AI companies. Though they were generally interested in personalized learning, there were still concerns about privacy:

\begin{quote}
    {\textit{``To customize to your skill levels, AI tools will be storing a lot of your personal data without telling you... And it can also be vulnerable to misuse... There's potential for it to also be leaked.'' --G3}}
\end{quote}

To address this, students discussed ways that AI tools could access personal data with protective limitations {(Access to Data, With Limitations)}. For example, G3's personalized learning tool ensured that private data used for training required clear consent and transparency, and that data would be retained only temporarily. Many students agreed with this method, suggesting data be \textit{``deleted after graduation unless the student says it's okay [to keep it]'' (G3)}, to safeguard privacy while enabling personalization.

\subsection{School AI Policy Designs}

In the second part, students developed school policies to mitigate specific harms of GenAI they had brainstormed. While students remained in their original small groups as described in section 3.2.2, some chose to focus on a drawback from a different problem area for their policy design. {Quotes in this section are taken directly from student policy discussions and presentations.}

\subsubsection{Policy 1: GenAI Cannot Be Used as a Legitimate Source}

{G1's} policy addressed students using inaccurate GenAI information {(AI Hallucinations)} in schoolwork by allowing GenAI to \textit{``help formulate or hone their ideas''} but prohibiting its use as a credible source in assignments. While the AI could suggest new sources, students must conduct the research themselves {(Balancing AI \& Learning)}, with the goal being that \textit{``students must assess the media they consume''}. Though this could make essay writing more time-consuming for students {and could lead to students utilizing powerful AI tools less}, it benefits students by ensuring they learn how to identify credible sources and gain valuable research skills. Additionally, it benefits teachers by informing them about \textit{``what the students struggled with, and what their skills are like truly.''}

\subsubsection{{Policy 2: }Limitations on Teacher Use of GenAI}

{G2} was concerned that teachers {(The Role of Teachers \& Parents)}, especially those with limited AI understanding {(Lack of Teacher AI Literacy)}, might over-rely on AI for student feedback. They proposed a policy requiring teachers to use standardized rubrics and spend a minimum amount of time on feedback to prevent this. While this could ensure better evaluations for students, it raised privacy concerns {(Access to Data, With Limitations)}, as teachers might feel uncomfortably watched---G2 noted they did not want \textit{``monitoring of teachers' computers during all hours''} but rather \textit{``their screen view when grading''} or \textit{``just logging chats [with the AI]''}. The group also felt more involved teacher feedback would benefit students and parents by potentially enhancing education quality.

\subsubsection{{Policy 3: }Require Parental Consent for GenAI Use}

G3 addressed issues around parental consent, as most high school students are under 18. Their policy required schools to obtain parental consent for AI use in the classroom, ensuring parents are aware of what their child is using in school and potential risks {(The Roles of Teachers \& Parents)}. However, G3 worried that withheld consent could create educational disparities {(Fair and Equitable Use of AI)}: \textit{``There’s going to be an issue if parents don’t consent. What would the student do? What would the school do to make sure that the student can have the same amount of education?''} They also worried about forcing teachers to manage unequal access to educational resources.

\subsubsection{Policy 4: GenAI Tools Must Delete Private Data}

G4 focused on using GenAI for personalized learning {(Personalized Learning)} while addressing data privacy risks. They worried students might feel \textit{``spied on''} by tools tracking data like text inputs or website access. To mitigate this, they proposed a policy requiring AI tools to delete student data when no longer needed {(Access to Data, With Limitations)}, such as after graduation or completing the relevant class, and limit the AI tool's access to certain websites to reduce the use of invasive anti-cheating measures {(Unreliable AI Detectors)}. Although this protects privacy, they felt it might reduce AI effectiveness through limiting training data and thus inadvertently encourage students to find other ways to cheat {(Academic Integrity)}. They also noted \textit{``teachers have an extra responsibility for checking through this platform''} {(The Role of Teachers \& Parents)} and that students would need to \textit{``learn how to use a new tool''}.

\subsection{{Survey Results}}

{Aiming to explore how students envisioned taking action to achieve their GenAI aspirations, one free-response} question of the post-survey asked students, \textbf{\textit{``How would you like to engage with generative AI tool or policy development going forward?''}}. {Five} themes were identified from responses: ``Help Administration'', ``Learn More'', ``Ethical Use'', ``Ensure Integrity'', and ``Casual Discussion''. {Most} students expressed interest in assisting school administration (``Help Administration'') on AI policy development, with responses {like} \textit{``maybe talk with superintendents,''} \textit{``discuss revising school policies with administration,''} or \textit{``bring the conversation to school deans''} {indicating students are interested in taking action by engaging in dialogue with authority figures.}

\begin{figure*}[ht!]
    \vspace{0pt} 
    \includegraphics[width=0.7\textwidth]{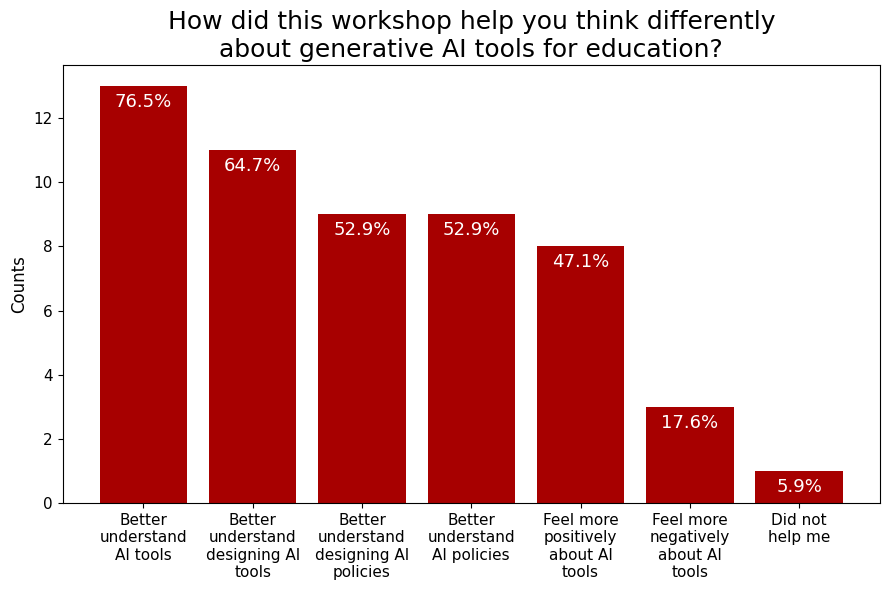}
    \caption{Results from the question ``How did this workshop help you think differently about generative AI tools for education?''}
    \Description{This figure shows a bar chart of results of the question ``How did this workshop help you think differently about generative AI tools for education?''.  13 responses or 76.5\% were ``Better understand AI tools'', 11 or 64.7\% were ``Better understand designing AI tools'', 1 or 5.9\% was ``Did not help me'', 9 or 52.9\% were ``Better understand designing AI policies'', 9 or 52.9\% were ``Better understand AI policies'', 3 or 17.6\% were ``Feel more negatively about AI tools'', and 8 or 47.1\% were ``Feel more positively about AI tools''.}
    \label{fig:diff-workshop}
\end{figure*}

Students were also asked the multiple-select question\textbf{\textit{ ``How did this workshop help you think differently about generative AI tools for education?''}}. Responses are in Figure \ref{fig:diff-workshop}, {and they demonstrate that the workshop effectively improved students' understanding of designing GenAI tools and policies. Notable results include that 13 students ({$76.5\%$}) felt the workshop improved their understanding of GenAI tools and 11 ({$64.7\%$}) felt it helped them better understand \textit{designing} GenAI tools}. Additionally, 11 {($64.7\%$)} students' views shifted either positively or negatively after the workshop. This may indicate students left with a deeper understanding of AI's benefits, increased awareness of AI's risks, and/or greater confidence in mitigating harms.

Finally, students were asked the multiple-select question, \textbf{\textit{``How do you feel about generative AI usage in schools?''}}. ``I feel knowledgeable about how generative AI can be used in schools'' was the most often selected option and had the most change from pre- to post-survey (from 7 selections or {$41.1\%$} to 15 selections or {$88.2\%$}, out of 17 total students), indicating that the workshop greatly improved students' understanding of AI's potential applications. 

\section{DISCUSSION}

In this section, we explore insights into student \textbf{challenges} and \textbf{needs} when integrating AI in schools, as well as proposed tools and policies to address these issues, answering RQ1. Regarding RQ2, we present \textbf{six guidelines} directly informed by student values and targeted at GenAI designers who wish to develop GenAI-based EdTech for high school environments. These guidelines bridge the gap between existing frameworks that provide higher-level advice and actionable steps that AI designers can take to ensure designs meet user needs. Finally, we call for future research and urge school administrations to involve student voices in developing school AI policies and rules, as students demonstrated a strong desire to contribute.

\subsection{{Design Insights from High School Student Challenges \& Needs}}

\subsubsection{Students Prefer System-facing Solutions}

Current literature predominantly advocates for educational approaches to AI safety, such as teaching privacy awareness \cite{huangarticle}, AI evaluation skills \cite{salazar2024generative, lin2021engaging}, and the ability to detect bias and misinformation in AI outputs \cite{ali2021children,lee2021developing}. These reflect digital literacy practices that place the burden on users to adapt to evolving technologies \cite{pangrazio2021towards}. In contrast, students in our workshop advocated for system-facing interventions targeting the design and governance of AI systems themselves, rather than relying solely on user education. For example, students advocated for diverse training datasets and external oversight to mitigate bias, aligning with emerging regulatory frameworks \cite{llms-yoon}. Students also emphasized embedding safeguards directly into AI systems, such as citing reliable sources and ensuring transparency \cite{daniella-policy,llms-yoon,codeorg2023}. They argued that these features would naturally equip users to handle misinformation \cite{el2024transparent}.

This highlights the need to shift from just teaching students to adapt to flawed AI systems to changing systems themselves. Our students, key stakeholders in EdTech \cite{zheng_charting_2024}, strongly favor implementing structural protections---like built-in citations, transparent decision-making, and trust-building mechanisms---over solely behavioral and educational adaptation. By emphasizing system-facing changes, they demonstrated their role as contributors to responsible AI design in education. These insights highlight the need for developers to prioritize transparency, equity, and accountability. Ultimately, they also illustrate that empowering users is more than equipping them to navigate flawed systems---it is about designing systems that empower users from the ground up.

\subsubsection{Accessibility Over Computational Power}

Accessibility concerns appeared consistently across all groups, reflecting broader discussions on digital equity \cite{liu2024digitalequity}, policy \cite{daniella-policy}, and how technology access disparities can exacerbate educational inequalities \cite{van2008technology}. Students emphasized designing AI tools that are compatible with various devices or accessible via web, ensuring equitable access regardless of available technology. These findings highlight students' capacity to advocate for equity within participatory design processes, as demonstrated in prior research \cite{ozer2020youth, martin2018iterative}. Our workshop expands this understanding to the context of GenAI, reaffirming students as critical stakeholders in creating equitable EdTech.

Notably, students prioritized accessibility over features like speed or computational power, advocating for GenAI tools to instead emphasize usability across a wide range of devices. This suggests that developers should focus first on creating inclusive, accessible tools---even if that requires performance trade-offs---to mitigate existing inequities.

\subsubsection{Academic Integrity Through Design}

While academic dishonesty remains a key concern with AI \cite{ventayen2023chatgpt,waltzer2023students}, students emphasized designing systems that naturally promote ethical use rather than focusing on detection and punishment. They highlighted unreliability of current AI detectors \cite{oravec2022ai, chaka2023detecting} and expressed concern that false accusations could undermine trust in teachers. Instead, students advocated for AI tools that encourage honest learning through design. They favored personalized hints over direct answers, consistent with prior research on AI's ability to promote learning through guided problem-solving \cite{maghsudi2021personalized}. This would allow them to develop AI proficiency---which they viewed as crucial for future careers \cite{alasadi2023generative, mesko2023prompt}---while maintaining essential skills \cite{buccinca2021trust,solyst2024children}. Based on student perspectives in the workshop, AI should be designed as a \textit{collaborative} partner in learning, balancing foundational skills with AI literacy and encouraging academic integrity through natural interaction. Key suggestions included progressive hint systems, reflection prompts, and transparent documentation of AI assistance---design choices that intuitively encourage ethical AI use.

\subsubsection{The Role of Teachers}

Throughout the workshop, students repeatedly emphasized the importance of teachers in the classroom while expressing concerns that limited AI knowledge among educators could lead to unintentional sharing of AI-generated misinformation. This aligns with prior research underscoring the need for teacher preparedness in AI adoption \cite{ayanwale2022teachers, vazhayil2019focusing, daniella-policy, codeorg2023}. Students also acknowledged challenges teachers face in building AI literacy due to demanding workloads \cite{collinson2001don}. By showing care for teachers, students demonstrated a compassionate approach as key stakeholders in AI integration \cite{zheng_charting_2024}. This suggests potential for collaborative learning, where students could support teachers’ AI literacy by co-learning or teaching concepts they understand \cite{fairman2004trading}. While not a substitute for professional development, this approach offers a temporary starting point to address knowledge gaps, fostering mutual growth.

\subsection{GenAI Design Guidelines for High School EdTech}

Drawing on insights from our participatory design process, we present guidelines for EdTech designers developing GenAI tools for high schools. We emphasize the importance of fostering discussions collaboratively with teachers, school boards, administrators, and---critically---high school students, who remain underrepresented in AI design conversations.
Existing frameworks for GenAI EdTech development, such as those from UNESCO and the DoE \cite{holmes2023guidance, doe_edtech_ai}, focus on education more broadly, rather than a specific age-group.
We focus on high schools, where students are more independent, seek greater control and understanding of AI \cite{chen2023integrating}, and have fewer policy efforts \cite{ghimire2024guidelines}, thus requiring more specific guidelines. High school is also a crucial transitory period \cite{blustein2000school, venezia2013transitions} and an important time for adolescents to develop critical thinking skills \cite{paul1989critical, schafersman1991introduction}. Furthermore, while existing guidelines emphasize foundational principles, we extend them with actionable strategies for high schools and future research directions.

\subsubsection{\textbf{Guideline 1: Prioritize Transparent and Consent-Based Data Practices}}
\hfill \break
\textit{Implement transparent data practices within GenAI tools that prioritize both educating students about data privacy and the implications of sharing personal information.} 
This approach enables a balanced exploration of the opportunities and risks associated with AI while fostering a commitment to regularly evaluating its use in schools. Existing guidelines \cite{holmes2023guidance} also highly value explicit consent from users.

Our findings revealed students were open to sharing personal data for improving the AI's contexual knowledge if they could control what is shared and ensure deletion after use. They balanced personalization with data governance, reflecting a nuanced understanding of the risks and benefits \cite{huangarticle}.
To address these, GenAI tools in high schools should include educational modules on data privacy that emphasize informed consent and user control over data. Students also should have the ability to manage data access, ensuring data is stored only with explicit permission and for the minimum necessary duration. Regular prompts about data privacy can further help students remain proactive in safeguarding their information. We also recommend future research investigation on forms of data management that high school students and teachers find most useful and intuitive.

\subsubsection{\textbf{Guideline 2: Foster Transparent AI Collaboration to Discourage Cheating and Build Trust}}
\hfill \break
\textit{Develop GenAI tools that actively involve students in maintaining academic integrity by promoting transparent and collaborative engagement.} Transparency, a principle emphasized in existing guidelines \cite{doe_edtech_ai, holmes2023guidance}, provides a foundation for building trust in AI systems.

Our study revealed significant student concerns about AI-facilitated cheating \cite{ventayen2023chatgpt} and stress caused by false accusations from AI detectors \cite{chaka2023detecting,oravec2022ai}. 
Students expressed apprehension about the potential inaccuracy of AI detectors and noted that peers might develop creative methods to evade these systems, exacerbating the issue.
In response, we propose transparent, collaborative GenAI tools that act as learning partners rather than surveillance mechanisms \cite{remian2019augmenting,akgun2022artificial}. They could provide real-time feedback regarding over-reliance or academic integrity concerns and invite students to explain thought processes, encouraging discussion. A ``self-disclose'' feature could help normalize ethical AI use by letting students share how they used AI. Additionally, transparent AI decision-making should give students access to the information AI detectors use to assess student work, easing concerns over false accusations. Teachers and students should establish clear communication about school policies surrounding cheating. This collaborative approach builds trust, positioning AI as a tool for learning, not policing. We also recommend future research with high schools to evaluate approaches that foster student trust in both AI systems and teachers, while ensuring the systems do not inadvertently enable cheating.

\subsubsection{\textbf{Guideline 3: Develop AI tools with Adaptive Accessibility Features}}
\hfill \break
\textit{Ensure GenAI tools are designed to adapt to both technological and socio-economic constraints, allowing for equitable access for all students and teachers.}  Accessibility is a value described in existing guidelines \cite{doe_edtech_ai, holmes2023guidance}, primarily in the context of supporting students with disabilities. However, this guideline extends this principle to address needs of all marginalized students with diminished access to technology.

Our findings revealed strong student support for GenAI tools with universal device accessibility, as all four student groups incorporated this feature into their designs.
We recommend GenAI tools to work seamlessly across various platforms, such as web browsers and devices with different computational capabilities, ensuring accessibility for all schools, regardless of resources \cite{liu2024digitalequity}. They should additionally function with minimal internet requirement to support equitable AI-enhanced learning, particularly for students in marginalized communities with limited technology and internet access outside school \cite{thomas2022closing}. Multilingual support is also crucial to prevent learning disparities \cite{alasadi2023generative} and bridge gaps between students from different socioeconomic backgrounds. Future research should focus on evaluating the performance of GenAI tools on lower-end devices and under limited networking conditions to ensure reliability.

\subsubsection{\textbf{Guideline 4: Prioritize Continuous AI Professional Development via Collaborative Teacher-Student Learning Programs}}
\hfill \break
\textit{Coordinate with schools to establish ongoing, adaptive professional development programs focused on teacher AI literacy. 
Incorporate a peer-to-peer learning model where digitally-savvy high schoolers serve as ``AI Ambassadors'' to collaborate with teachers and support ongoing professional development. } Educator AI literacy is emphasized in existing guidelines \cite{doe_edtech_ai, holmes2023guidance}, though they offer limited concrete strategies for its improvement.

Studies show teachers feel unprepared for GenAI integration and desire more structured training \cite{ding2024enhancing, langreo2023ai}. Students in our study stressed that successful GenAI use relies on teachers’ AI literacy, expressing concern that this need could further burden teachers unless adequately supported by schools \cite{collinson2001don}.
Designers should advocate for and help create school-sponsored professional development, providing teachers with GenAI knowledge and practical classroom strategies through hands-on, scenario-based training tailored to different skill levels. They should regularly update trainings as technology improves and receive regular feedback. A peer-to-peer model where tech-savvy students act as co-educators, or ``AI Ambassadors'', can also help teachers understand student perspectives, reduce workload, and build student trust in teachers' AI skills. Collaborative AI integration alleviates pressure on teachers to independently tackle AI literacy and fosters a supportive community. Researchers should work with students and teachers to develop and evaluate these peer-to-peer learning curricula for comprehensive coverage.

\subsubsection{\textbf{Guideline 5: Foster Responsible AI Integration in Education}}
\hfill \break
\textit{Design AI systems that promote a balanced approach to AI for learning, enhancing efficiency while encouraging development of core skills and independent problem solving. } Existing guidelines \cite{holmes2023guidance} reference the importance of foundational skills and preventing over-reliance, but do not provide actionable steps.

Our results revealed students were concerned about potential over-reliance on AI, leading to a decline in foundational skills \cite{buccinca2021trust}.
They advocated for a balanced approach, believing that AI proficiency is vital for future careers.
For this, educational AI systems should function as supportive assistants \textit{and} mentors, guiding students to use foundational skills before relying on AI. Instead of offering direct answers, these tools should inspire students to articulate thought processes, work through core competency exercises, and develop problem-solving strategies. This could be enhanced with reflective practices that prompt students to assess how AI influenced their approach, what insights they gained, and how their understanding evolved. These exercises promote independence and critical thinking, helping students develop a balanced skill set that combines traditional learning with important AI competencies, such as prompt engineering \cite{mesko2023prompt}. More research is needed to explore the impact of AI integration and identify AI-guided learning approaches that maximize student learning while minimizing unhealthy dependence.

\subsubsection{\textbf{Guideline 6: Implement Robust AI Regulation and Accountability Mechanisms in Schools}}
\hfill \break
\textit{Collaborate with schools and policymakers to establish a ``Transparent AI Certification'' program to ensure educational GenAI tools meet high, third party standards for fairness, accuracy, and accountability. } Existing guidelines \cite{doe_edtech_ai} focus on national regulation and ethical standards but offer less on actionable frameworks or local regulation to ensure AI tools meet enforceable standards.

Our research shows students want AI regulation but are unsure about the best approach. They suggested options involving the government or third-party companies, and wanted to decentralize control. They also feared school-level regulation might not effectively address systemic issues like bias or misinformation and could lead to excessive monitoring \cite{akgun2022artificial}. As students were unclear about exact types of regulation but favored external oversight, our recommendation focuses on actionable regulation without specifying at what level.

A ``Transparent AI Certification'' program should be developed that requires AI tools to meet criteria set by a trusted third party for transparency, bias mitigation, and accuracy. Developers can contribute to the design of robust certification guidelines and their benchmarks in partnership with independent regulatory bodies, government agencies, educators, students, and administrators. They should also provide guidance to help schools adapt the program to their individual needs. The certification should evaluate AI tools' abilities to provide unbiased, accurate information and require features like real-time fact-checking. Developers should additionally commit to performing ongoing, third-party audits of their tools and update their tools to reflect changes in best practices. Integrating an ``AI Accountability Dashboard'' as part of the certification can present information on the AI’s decision-making processes, data sources, and any constraints in its use, similar to suggestions from previous policy reports \cite{llms-yoon}. 
Future research is needed to explore student and teacher understanding of the ``Transparent AI Certification'' and ``AI Accountability Dashboard'' as these functionalities are only effective when fully understood by users.

\subsection{Giving Students a Voice in School AI Policies}

The guidelines presented in 5.2 focus primarily on actionable strategies for AI designers, some of which require collaboration with schools. In this section, we turn our attention to how schools can independently develop effective and understandable school AI policies based on student perspectives from our workshop. 

We urge schools to formally \textbf{involve students in development of school AI policies}, ensuring their voices are reflected in decision-making. Students, as primary stakeholders, offer valuable insights and a strong desire to contribute to school policies, ensuring that AI policies are influenced by those most impacted---students themselves---making it a crucial addition to existing frameworks. Our findings reveal that students see themselves as key participants in school AI policy discussions and are particularly interested in collaborating with administrators. Survey responses suggested improved student understanding of AI development post-workshop, with continued involvement likely to further enhance interest and engagement, as supported by participatory design research \cite{bell2016learning,triantafyllakos2008we,wu_investigating_2021}. Additionally, student involvement in decision-making boosts engagement and connection with technologies in their environment \cite{levin2000putting, dindler2020computational}.

To achieve this, schools should establish student-led AI advisory committees that promote collaboration on school policy development between students, teachers, and administrators. Schools should also provide platforms for anonymous student feedback on AI tools and school policies. Integrating AI policy discussions into curricula would additionally prepare students to contribute to school AI policy development---and AI policy in general---and involving students in pilots for new AI tools would ensure they can better inform school-level decisions. This approach acknowledges students as key stakeholders, promoting a culture where primary users of technology have a voice \cite{bjorgvinsson2010participatory,tuhkala2021systematic}. Prior research also emphasizes the importance of clear rules and transparent dialogue in promoting effective learning \cite{marzano2003key, purkey1983effective}. Regular communication of school AI policies and the reasoning behind them---particularly when student voices are included---builds trust between students and educators, promoting a shared understanding of how technologies uphold school values \cite{gillespie2005student}. 

Finally, we recommend future research and periodic audits to assess the extent to which school policies incorporate student input and how effectively they align with student needs and perspectives.

\section{LIMITATIONS \& FUTURE WORK}
Our recruitment method of interest-based mailing lists likely attracted participants with pre-existing interest in GenAI. Additionally, our U.S.-based sample included 58.8\% private school participants, compared to the 2021 national average of 18.1\% \cite{fabina_school_2023}. While this sample provided valuable insights, future research would benefit from recruiting students without prior GenAI interest and including a more diverse range of school types. As with all qualitative research, while our sample size ($N=17$) and demographic composition limit generalizability, our goal was not to produce universally applicable findings but rather to provide deep, nuanced insights into how our high school participants perceive and engage with GenAI in their education.

Extending participatory design workshops over longer periods could also offer deeper insights into students' evolving views on AI in education. Incorporating iterative AI tool design with tangible prototypes and teacher feedback would help provide more concrete strategies for effectively implementing GenAI in classrooms.

While we aimed to provide a unified understanding of GenAI for workshop participants, we did not formally test this knowledge beyond student discussion analysis. Variations in understanding and personal experience may have influenced tool and school policy designs, presenting another avenue for future investigation.

Future research should also explore both the practical implementation of our proposed guidelines in diverse school settings and the potential challenges that may arise. Additionally, it is important to examine how these guidelines may impact equity, particularly with regard to access and outcomes for historically underrepresented student groups.

\section{CONCLUSION}

This paper presents insights from high school students---a demographic often overlooked in participatory design of EdTech---regarding GenAI tools and school policies. Through a participatory design workshop, students described their needs and challenges surrounding GenAI, designing GenAI tools and school policies to mitigate the issues they found most pressing: cheating, plagiarism, misinformation, and over-reliance on AI. Based on their perspectives, we proposed six actionable guidelines for GenAI developers targeting EdTech for high schools, emphasizing transparency, trust, accessibility, and balanced integration. Additionally, students' strong interest in contributing to school AI policies suggests untapped potential for their involvement in school AI policy development. 

\section{SELECTION AND PARTICIPATION OF CHILDREN}
All elements of this research study were approved by our university’s Institutional
Review Board. Additionally, all facilitators of our workshop went through appropriate ethics and safety training. Recruitment of children took place through interest-based mailing lists about AI and robotics. Participants voluntarily indicated their interest through an online form, and all interested students were invited to participate in the workshop. We recruited 17 high schoolers between the age of 14 and 17, with an average age of 16, who attended public ($N=7$) and private ($N=10$) schools. Students were provided with assent forms and parents were provided with consent forms informing them about the workshop content and asking for their agreement to collect and disseminate data for scientific purposes. Parents and students were also informed about the goals of the study, any potential safety and privacy risks, and how we would protect student data.


\bibliographystyle{ACM-Reference-Format}
\bibliography{hs-student-teacher-codesign}

\appendix

\end{document}